\documentstyle[preprint,aps,prb,amsfonts]{revtex}

\begin{document}

\draft

\title{
 Free induction signal from biexcitons and bound excitons}
\author{Emmanuel I. Rashba}
\address{
Department of Physics, University of Utah, Salt Lake City, Utah 84112 \\
and L. D. Landau Institute for Theoretical Physics, Moscow 117940, Russia}
\date{23 December 1996}

\maketitle


\begin{abstract}
 A theory of the free induction signal from biexcitons and bound excitons
 is presented. The simultaneous
 existence of the exciton continuum and a bound state is shown to
 result in a new type of time dependence of the free induction. 
  The optically detected signal increases in time and oscillates with
 increasing amplitude until damped by radiative or dephasing processes.
 Radiative decay is anomalously fast and can result in strong
 picosecond pulses.
 The expanding area of a coherent exciton polarization (inflating antenna),
 produced by the exciting pulse, is the underlying physical  mechanism.
 The developed formalism can be applied to different biexciton transients.

\end{abstract}


\section{Introduction}
\label{sec:introduction}

Ultrafast spectroscopy of excitons in the time domain\cite{fast}
has proven to be a powerful tool to probe quantum coherence of
 exciton states, which was originally studied by polarization 
 of the stationary emission.\cite{PIv} Most of the experimental data
 were taken from GaAs quantum wells, however, some experiments were
 performed with bulk excitons.  Quantum beats in
 different response functions provide a manifestation
 of the coherence driven by external fields. These  beats appear when 
 several states having close energies are excited simultaneously.
 Quantum beats were observed with magnetically split exciton
 levels,\cite{Osten} bound 
 excitons with different confinement energies,\cite{Gobel}
 heavy and light hole excitons,\cite{Feuer} free and bound excitons,\cite{Leo}
 and with biexcitons and a two-exciton continuum.\cite{biex,WSD} Two last
 systems have much in common since they possess both a continuous
 spectrum and a discrete level. They will be of principal 
 importance for us in what follows.

 Experimental data provide convincing evidence of a strong effect
 of the exciton-exciton and exciton-free-carrier interaction on nonlinear
 response functions.\cite{resp,Wang} However, there could not be universal
 approach to the theoretical treatment of this nonlinear many-electron
 problem. The physical patterns are rather different in different
 ranges of the parameter values, therefore, the theoretical approaches
 also should be different.
 When the intensity of a pump is high and the nonradiative
 dephasing time $\tau$ is short, the Hartree-Fock approximation works
 rather well. This approach was advanced by Schmitt-Rink and Chemla\cite{S-RC}
 and has been  developed in a number  of papers.  It is based on the 
 semiconductor Bloch equations\cite{S-RCH,LBK} and allows a generalization
 accounting for two-exciton correlations.\cite{Sham}
 Time evolution of wave packets produced by the pump also can be
 followed.\cite{Glut} It is a distinctive feature of this approach
 that the basis of the exciton and electron-hole pair states is restricted
 by excitations strongly coupled to the pump, 
  i.e., possessing the momenta
 of the exciting electromagnetic waves.
 This restriction is the basic limitation of the powerful Hartree-Fock method.
 A different approach is  used in the opposite limit of
 low exciton densities, when the
 interaction of biexcitons can be neglected and the dephasing time $\tau$
 is long. In this limit biexcitons can be considered as noninteracting
 particles, and the problem can be reduced to the dynamics of an exciton
 molecule in external fields. Even a simple four-level biexciton
 energy scheme\cite{FinSai,Bott} turned out to be very successful. In the
 framework of this scheme the biexciton continuum is substituted by a
 single two-exciton state. The four-level scheme
 allowed one to explain the existence of beats, having a period of inverse
  biexciton binding energy,  in different nonlinear phenomena.
 More recently,  Wang {\it et al.}\cite{WSD} have discussed the effect
 of the biexciton continuum on initial biexciton transients. These
 authors neglected the exciton-exciton interaction in the biexciton continuum
 and investigated quantum evolution of the relative momentum of two
 excitons in a molecule. Our paper is related to the approach 
 advanced in Refs.~\onlinecite{FinSai,Bott}, and \onlinecite{WSD}.
 Developing a theory that allows for a consistent account of the 
 biexciton continuum and the exciton-exciton interaction is the main
 objective of the  paper. We show that the interaction of excitons 
 and their dynamics are of critical importance for biexciton transients.
 Our results are primary developed for an exciton in the field of a defect.
 This problem is of intrinsic importance and is actively studied. Fast
 early-time evolution of the Rayleigh signal from inhomogeneously
 broadened exciton levels was recently investigated both theoretically
 and experimentally.\cite{ZCH}

 The traditional approach to quantum beats is based on the energy spectrum 
 comprising few, usually two, discrete energy levels. This approach can be
 applied to  beats between heavy- and light-hole excitons
 because of the momentum 
 conservation and absence of the interaction between these excitons.
 However, the biexciton and bound exciton problems are more involved.
 Indeed, energy spectra of these systems include the two-exciton and 
 single-exciton
 continuum, respectively. For example, for a two-exciton system it is
 the exciton-exciton
 interaction that supports the two-photon coherence, and 
 the lower part of the continuum with the width of about several biexciton
 binding energies, $\varepsilon_b$, contributes to the coherent
 polarization along with the bound biexciton state.
 In addition to the theoretical arguments, the
 experimental data provide weighty, although indirect, evidence
 of the role of the two-exciton continuum.
 Indeed, it was shown\cite{FinSai}
 that the four-level biexciton energy scheme  can be brought
 into agreement with experimental data only if the  enhancement factor
 typical of giant oscillator strengths\cite{GOS,GR73,Han} is invoked.
 Therefore, frequencies of allowed transitions are distributed
 continuously and the spread of the frequencies is about 
 $\varepsilon_b$. Naive consideration suggests that
 such an energy spectrum should result in beats having a frequency of
 about $\varepsilon_b$ and showing fast nondissipative decay
 because of accumulating phase differences between different modes.
 It turns out that the actual physical picture is quite different.

  In this paper we present an exact solution for a free induction signal
 excited by a single-side exponential pulse, $t<0$, in a nondissipative system
 with a biexciton nonlinearity. This signal can be observed, e.g., as
 a resonance fluorescence at positive times, $t>0$.
The special shape of the pulse
 simplifies calculations but does not influence the basic results.
 The contribution of the two-exciton
 continuum is consistently taken into account.\cite{recent}
 With such an approach, free induction, i.e., free oscillations
 of a two-exciton wave function $\Psi(t)$
 for $t>0$, includes two modes. There exists {\it a
 beating mode} describing {\it undamped beats whose frequency is
 equal to} $\varepsilon_b$. Damping of this mode is controlled by
 the mechanisms other than the exciton dispersion 
 (radiative decay, dephasing, polariton effects, etc.).
 There exists also {\it a growing mode} whose amplitude increases linearly
 with time $t$ and whose carrier frequency
 equals the energy of the bottom of the
 two-exciton continuum. The growing mode is inherent in interacting
 systems possessing a continuous spectrum. It is the inflation
 in a real space of the wave packet created by the pulse
 that manifests itself in this mode. The packet is termed 
 an {\it inflating antenna} in what follows. The two modes result in 
 an optically detected free induction signal that (i) increases with $t$,
 (ii) has monotonic and oscillating parts, and (iii) results in ultrafast
 radiative decay. The same
 modes exist for excitons bound to impurities. Both  modes originate
 from the analytical structure of the two-exciton Green function,
 which is specific for Hamiltonians with a continuous spectrum and
 violated momentum conservation.
 We expect that these modes contribute also to different nonlinear processes,
 including multiple-impulse processes, and that the developed
 technique is of general applicability.

\section{Bound excitons: General formalism}
\label{sec:bound}

 To make clear the basic idea and account rigorously for analytical
 properties of the exciton Green functions, we develop an exactly
 soluble model.
 To this end we neglect polariton effects,\cite{polar} which will
 be briefly discussed in Sec.~\ref{sec:Disc}, and dephasing. We also neglect
 the dependence of the scattering amplitude on the light polarization
 because it is sensitive
 to the band structure, geometry, etc.\cite{Wang,CWS}
 Excitons are considered stable particles without internal
 degrees of freedom. It is convenient
 to start with the bound exciton problem. The free induction signal from bound
 excitons can be found in the framework of the linear response approach.
 Therefore, the specific spectral properties of the inflating antenna
 can be completely disentangled from  nonlinear phenomena.

 If the electromagnetic wave,
\begin{equation}
 {\bf E}_{\bf q}({\bf r}, t)
 \propto \exp\{i({\bf q}\cdot{\bf r} - \omega t)+\alpha t\}~,~~
 ~\alpha>0~,
\label{eq0}
\end{equation}
 is incident upon a crystal at $-\infty < t < 0$, the
 exciton wave function $\Psi_{\bf q}({\bf q}', t=0)$
 can be calculated as a linear response to this perturbation.
 Free evolution of $\Psi_{\bf q}({\bf q}', t=0)$ determines the wave
 function $\Psi_{\bf q}({\bf q}', t)$ at any instant $t$, $t>0$:
\begin{equation}
 \Psi_{\bf q}({\bf q}', t) = (iM/{\sqrt v} )
 A_{{\bf q}'{\bf q}}(t)~, ~~ 
 A_{{\bf q}'{\bf q}}(t) = \sum_{j}
 {{ \psi_{j}({\bf q}'){\bar \psi}_{j}(\bf q)}
 \over {\omega-E_{j}+i\alpha}} {\rm e}^{-iE_{j}t}~.
\label{eq1}
\end{equation}
 Here  $\Psi_{\bf q}({\bf q}', t)$ and $\psi_{j}({\bf q}')$
 are, respectively, time-dependent and stationary exciton wave functions
 in the momentum representation. The subscript
 $j\geq 1$ numerates single-exciton states, bound and free ones.
  The ground-state energy of the
crystal is chosen as the origin, $E_{0}=0$; $\hbar =1$.
  The coefficient $M$ is
 the matrix element, per unit cell, of  the perturbation produced by
 the field ${\bf E}_{\bf q}({\bf r}, t)$, and $v$ is the unit cell volume. 
 The momenta $\bf q$
 and ${\bf q}'$, which are of importance for optical experiments,
 are small and will be neglected in the final results. Equation (\ref{eq1})
 does not take into account the radiative decay of the wave packet.
 The role of this process will be discussed in Sec.~\ref{sec:induction} below.

 The amplitude $A_{{\bf q}'{\bf q}}(t)$ describes free precession of
 $\Psi_{\bf q}({\bf q}', t)$ for positive times.
 For $t=0$, $A_{{\bf q}'{\bf q}}(t=0)$
 coincides with the retarded exciton Green function
\begin{equation}
 G_{{\bf q}'{\bf q}}(\omega) = \sum_{j}
 { \psi_{j}({\bf q}'){\bar \psi}_{j}(\bf q)}/(\omega-E_{j}+i0)
\label{eq2}
\end{equation}
 if the argument $\omega$ is substituted by $\omega + i\alpha$.
 For arbitraty $t$, functions
 $A_{{\bf q}'{\bf q}}(t)$ and $G_{{\bf q}'{\bf q}}(\omega)$
 are related by the equation
\begin{equation}
 A_{{\bf q}'{\bf q}}(t) =
 {  1\over {2\pi i}}
 \int_{-\infty}^{\infty} d\omega'~
 { \exp(-i\omega't) \over {\omega' - \omega - i\alpha}}
 G_{{\bf q}'{\bf q}}(\omega')~.
\label{eq3}
\end{equation}
Equation (\ref{eq3}) can be checked by employing Eq.~(\ref{eq2}), closing
 the integration path in the lower complex half-plane, and calculating
 residues in the poles of $G_{{\bf q}'{\bf q}}(\omega')$.

 Subsequent transformations of $A_{{\bf q}'{\bf q}}(t)$
 are based on the introduction of the scattering operator\cite{Baz}
 ${\hat {\cal T}}(\omega)$ 
\begin{equation}
 G_{{\bf q}'{\bf q}}(\omega) =
 G_{\bf q}^{0}(\omega) \delta_{{\bf q}'{\bf q}} +
 G_{{\bf q}'}^{0}(\omega)
 {\cal T}_{{\bf q}'{\bf q}}(\omega) G_{\bf q}^{0}(\omega)~,
\label{eq4}
\end{equation}
 where $G_{\bf q}^{0}(\omega)=[\omega - \varepsilon({\bf q})+i0]^{-1}$
 is a free-exciton Green function.
 Only the second term of Eq.~(\ref{eq4})
 contributes to $A_{{\bf q}'{\bf q}}(t)$ for
 ${\bf q}'\neq{\bf q}$ and will be retained below. It is an important
 property of this term that it includes a product of two $G^0$ functions.
 Their poles nearly coincide for ${\bf q}\approx {\bf q}'$.
 This property strongly influences the
 subsequent results. It is convenient to split the scattering operator
 ${\hat {\cal T}}$ into two terms as
\begin{equation}
 {\hat {\cal T}}({\omega}) = {\hat T}_{\infty} + {\hat T}({\omega}) ~,
\label{eq4a}
\end{equation}
 where ${\hat T}_{\infty} = {\hat {\cal T}}({\omega}={\infty})$. 
 The function $T_{{\bf q}{\bf q}'}(\omega)$ 
 is analytical in the upper half-plane,
 hence, the following Lehmann representation\cite{AGD} holds for it
 \begin{equation}
{\hat T}(\omega) = - {1\over \pi} 
 \int_{-\infty}^{\infty} d\omega'~{\hat T}^{''}(\omega')/
 (\omega - \omega' + i0)~,
\label{eq5}
\end{equation}
 where ${\hat T}^{''}(\omega) = {\rm Im}\{{\hat T}(\omega)\}$. Sustituting
 Eqs. (\ref{eq4}) - (\ref{eq5}) into Eq. (\ref{eq3}) and performing integration
 over $\omega'$ by closing the integration path in the lower
 half-plane, one gets after some algebra
\begin{eqnarray}
 &&A_{{\bf q}'{\bf q}}(t) =
 {1\over {\varepsilon({\bf q}') - \varepsilon({\bf q})}} 
 \biggl \{ ({\hat T}_{\infty})_{{\bf q}' {\bf q}}
 \biggl [{\exp(-i\varepsilon ({\bf q}')t)\over 
 {\omega - \varepsilon ({\bf q}') + i\alpha}}
 - {\exp(-i\varepsilon ({\bf q})t)\over
 {\omega - \varepsilon ({\bf q}) + i\alpha}} \biggr ] - {1\over \pi}
 \int_{-\infty}^{\infty} dx~ T^{''}_{{\bf q}'{\bf q}}(x)\nonumber \\ &\times&
  \biggl[{1\over {x-\varepsilon({\bf q}')}}
  \biggl( {\exp(-ixt)\over {\omega-x+i\alpha}} -
 { \exp(-i\varepsilon({\bf q}')t)\over
 {\omega - \varepsilon({\bf q}')+i\alpha}}\biggr)   
  - 
 {1\over {x-\varepsilon({\bf q})}}
 \biggl( {\exp(-ixt)\over {\omega-x+i\alpha}} -
 { \exp(-i\varepsilon({\bf q})t)\over
 {\omega - \varepsilon({\bf q})+i\alpha}}\biggr) \biggr ] \biggr \}~.
\label{eq6}
\end{eqnarray}
 Equation~(\ref{eq6}) provides an exact expression for the exciton wave function
 in the momentum representation. The wave function in the coordinate
 representation can be found by the Fourier transformation in the variable
 ${\bf q}'$. This transformation can be performed only numerically and
 depends on the exciton dispersion law and the interaction potential
 between the exciton and the impurity. However, for optical applications
 this transformation is not needed. The photon scattering amplitude
 is completely determined by the amplitude $A_{{\bf q}'{\bf q}}(t)$
 in the small ${\bf q}$ and ${\bf q}'$ region. Under these conditions the
 energies $\varepsilon({\bf q})$ and $\varepsilon({\bf q}')$ are nearly equal.
 For $ {\bf q}', {\bf q}\rightarrow 0$ both the numerator and denominator
 vanish and Eq.~(\ref{eq6}) takes the form 
 \begin{eqnarray}
 A(t) &=& T_{\infty} ~ {d\over {d\varepsilon}} 
 \biggl({ \exp(-i\varepsilon t)\over
 {\omega - \varepsilon + i\alpha}} \biggr) \nonumber \\
 &-& {1\over \pi} ~
  \int_{-\infty}^{\infty}d\omega' ~
  T^{''}(\omega') ~ {d\over {d\varepsilon}} 
 \biggl[{1\over {\omega'-\varepsilon}}
 \biggl( 
 {\exp(-i\omega't)\over{\omega-\omega'+i\alpha}}
 - {\exp(-i\varepsilon t)\over {\omega -\varepsilon +i\alpha}}
 \biggr) \biggr]~.
\label{eq7}
\end{eqnarray}
 Here $\varepsilon = \varepsilon(0)$
  is the energy of long-wave-length excitons,
 and $A(t)$, $T_{\infty}$,  and $T^{''}(\omega')$ are the limits of
 $A_{{\bf q}'{\bf q}}(t)$,  $ (T_{\infty})_{{\bf q}'{\bf q}}$~,
 and $T_{{\bf q}'{\bf q}}(\omega)$,
 respectively, for ${\bf q}, {\bf q}'\rightarrow 0$. Equation (\ref{eq7})
 is the final equation for the time-dependent amplitude $A(t)$.
 It is completely determined by $T_{\infty}$ and
 the imaginary part of the operator ${\hat T}(\omega)$.

\section{Free induction signal}
\label{sec:induction}

 Equation (\ref{eq7}) determines the time dependence of the free induction
 signal. Two basic properties of this equation follow from general arguments.

 First, the derivatives $d/d\varepsilon$ result in a contribution to
 $A(t)$ proportional to $t\exp(-i\varepsilon t)$. 
 This property is obvious as applied to the first and third terms
 of Eq.~(\ref{eq7}). A similar contribution from the second term can be
 found by using the identity $d(\omega'-\varepsilon)^{-1}/d\varepsilon =
 - d(\omega'-\varepsilon)^{-1}/d\omega'$ and transforming the integral
 by parts. The term in the amplitude $A(t)$ increasing with $t$ will be
 termed below as {\it the growing mode}.\cite{log}
 It originates from the product of two $G^0$
 functions with nearly coinciding poles, Eq.~(\ref{eq4}).
 The growing mode is reminiscent
 of the growing solutions of differential equations with degenerate
 characteristic numbers. This mode describes the
 global evolution of the wave
 packet prepared by the pulse. As $t$ increases, the packet
 expands in  $\bf r$ space. For translationally invariant systems this
 expansion is accompanied by changes in the phases of different Fourier
 components, whereas their moduli remain unchanged. The
 impurity potential, attractive or repulsive,
 violates the momentum conservation. Consequently,
 the amplitude of the ${\bf q}=0$ mode grows. It indicates that
 the volume of the coherence area around the impurity  
 increases with $t$. The giant oscillator strengths observed
 in stationary experiments were ascribed by Thomas and Hopfield
 to {\it exciton antennas}.\cite{TH}
 In these terms the growing mode is generated by an
 {\it inflating exciton antenna}. This picture explains why the growing mode
 is specific for systems possessing a continuous spectrum and,
 hence, extended states. Mathematically this mode
 originates from the integration of $G_{{\bf q}' {\bf q}}(\omega)$ 
 along the cut
 in the complex plane passing across the exciton band.

 Second, the bound state is a pole of~ ${\hat T}(\omega)$. Therefore,
 $T^{''}(\omega')$ includes a term proportional to 
 $\delta(\omega' - \varepsilon + \varepsilon_{b})$, where $\varepsilon_{b}$
 is the binding energy of an exciton to the impurity center.
 This term contributes into Eq.~(\ref{eq7}) 
 an oscillating exponent $\exp[-i(\varepsilon -\varepsilon_{b})t]$.
 The exciton bandwidth is supposed to be large as compared 
 with $\varepsilon_b$. Under these conditions the integration along the cut in
  the complex plane 
  contributes the factor $\exp(-i\varepsilon t)$.
 Two oscillating terms in Eq.~(\ref{eq7}),
 $\exp[-i(\varepsilon -\varepsilon_{b})t]$ and
 $\exp(-i\varepsilon t)$, result in  beats at the frequency of
 $\varepsilon_{b}$ with a time-independent amplitude. It is remarkable
 that the oscillations remain undamped despite the fact that
 $\varepsilon$ belongs to a continuous spectrum.
 The scillating contribution to $A(\omega)$ will be termed {\it the
 beating mode}.

 The scattering operator ${\hat {\cal T}}(\omega)$
 can be easily found for a Frenkel exciton
 when the impurity potential is described by a degenerate perturbation
 $U_{{\bf m}{\bf n}} =  U \delta_{{\bf m}0}\delta_{0{\bf n}}$,
 $U < 0$.\cite{BRS}
 Here $\bf m$ and $\bf n$  denote sites, and the impurity
 resides at the site ${\bf m}={\bf n}=0$. Under these conditions
 ${\cal T}_{{\bf q}' {\bf q}}(\omega )$ does not depend on the momenta
 ${\bf q}$ and ${\bf q}'$ and equals
\begin{equation}
 {\cal T}(\omega) = {U\over {1-UG_{0}(\omega)}}~, ~~
 G_{0}(\omega) = \int{\rho(\omega ')\over {\omega - \omega ' + i0}}
 ~d\omega '~,
\label{eq8.0}
\end{equation}
where $G_{0}(\omega)$  denotes
 the free Green function $G^{0}_{\bf q}(\omega)$
 integrated over the momentum $\bf q$, and $\rho(\omega)$ is the
 density of states inside the exciton band. It is easily seen that
  $T_{\infty} = U$.  Binding energy
 $\varepsilon_{b}$ is related to the potential $U$ by the
 equation $G_{0}(\varepsilon - \varepsilon_{b}) = U^{-1}$.
 For a  two-dimensional (2D) system the density $\rho(\omega)$ can
 be modeled as $\rho(\omega)=1/E_B$ for $0 \leq \omega \leq E_B$~,
 where $E_B$ is the bandwidth, and the
 integral in Eq.~(\ref{eq7}) can be performed. Finally
\begin{equation}
 {\cal T}(\omega) = E_{B}/ {\biggl\{ \ln \biggl[
 { \varepsilon_{b} (E_{B} - \omega)\over
 {\omega (E_{B} + \varepsilon_{b}) }}\biggr]
 + i\pi \biggr\}}~.
\label{eq8}
\end{equation}

  The amplitude $A(t)$ is shown in Fig.~\ref{fig1}
 for three values of $\omega$.
 Both the linear-in-$t$ growth and the
 oscillations with a time-independent amplitude are distinctly seen
 in the asymptotic region $t \agt 2\pi \varepsilon_{b}^{-1}$.  Actually,
 they are seen even for small values of $t>0$, but the shape of the
 first oscillation is somewhat distorted. 
 The data for a 3D system with a model density
 $\rho(\omega) = 8\sqrt{\omega (E_{B}-\omega)}/\pi {E_{B}}^2$ are also
 shown in Fig. 1. One can see that the dependence of $A(t)$ on dimensionality
 is rather weak. The linear-in-$t$ growth of $A(t)$ in the large $t$ region
 originates from the exciton dynamics. It disappears in the 
 $E_{B}\rightarrow 0$ limit when the exciton effective mass tends
 to infinity.
 
 Therefore, after a short transient  the growing and beating modes
 dominate the amplitude $A(t)$. The optically detected
 free induction signal $I(t)$ is related to the zero-momentum
 component of the wave function, $I(t)\propto |\Psi_{0}(0, t)|^2$.
 In the asymptotic region
\begin{equation}
 A(t) \propto \biggl\{t + {b\over 2} \exp[i(\varepsilon_{b}t + \phi)]\biggr\}
 {\rm e}^{-i\varepsilon t}~,
\label{eq9.0}
\end{equation}
 where $b$ and $\phi$ are real parameters, and
 $I(t)$ obeys the following law:
\begin{equation}
 I(t) \propto \{t^2 + bt \cos(\varepsilon_{b}t + \phi)\}~.
\label{eq9}
\end{equation}
 Interference of the two modes
 results in an unusual shape of the signal $I(t)$.
 It consists of the monotonic and oscillatory contributions, and both of them
 growing (rather than decaying) with $t$.\cite{note}

 Figure \ref{fig1} displays the typical shape of the response $A(t)$.
 It is instructive to consider also a special case when a spectrally
 narrow pulse is in resonance with the bound-exciton state. The data
 are shown in Fig.~\ref{fig2}. For $t\ll \alpha^{-1}$ the
 amplitude $|A(t)|$ is nearly constant and describes a bound exciton.
 As $t$ increases, the admixture of the continuum grows and the
 shape of the amplitude described by Eq.~(\ref{eq9.0}), linear growth
 and strong undamped oscillations, sets in for 
 $t\agt 3{\alpha}^{-1}$. If the pulse is out of
 resonance with the bound exciton, the shape of the response similar
 to that of Fig.~\ref{fig1} is recovered.

 The above theory does not take into account the radiative decay
 of the exciton wave packet. It is this approximation  that results
 in the unlimited increase in the amplitude $A(t)$, Eq.~(\ref{eq9.0}).
 If one neglects the oscillating part of Eq.~(\ref{eq9}), 
 the radiative lifetime 
$\tau_{R}(t)$ decreases with $t$ as $\tau_{R}(t)\propto t^{-2}$.
 This rapid increase in the emission probability implies a non-Lorentzian
 shape of the emission signal and establishes the applicability limit for
 Eq.~(\ref{eq9}). A rigorous approach should include a generalization
 of the Weisskopf-Wigner theory as applied to the exciton antenna.
 We restrict ourselves to a phenomenological approach, which allows us
 to estimate the duration of the emitted pulse. 

 One can infer from Fig.~\ref{fig1} that if $\alpha \sim \varepsilon_b$
 and the carrier frequency $\omega$ of the exciting pulse is not far from
 the exciton band bottom
 $\varepsilon$, the radiative lifetime $\tau_{R}(t)$ can
 be evaluated as
\begin{equation}
\tau_{R}^{-1}(t) \approx (\tau_{R}^{0})^{-1} (1 + \beta \varepsilon_{b} t)^2 ~,
\label{eq9.1}
\end{equation}
 where the coefficient $\beta \sim 1$.
 The wave packet size at $t=0$ is approximately equal to 
 the bound-exciton radius, hence,  $\tau_{R}^{0}$ can be estimated
 as the bound-exciton radiative lifetime. Since a single light
 quantum should be radiated in the emitted pulse, the duration
 $\tau_{\rm em}$ of this pulse can be estimated from the equation
\begin{equation}
\int_{0}^{1} dn = \int_{0}^{\tau_{\rm em}} dt/{\tau_{R}(t)} = 1~.
\label{eq9.2}
\end{equation}
 Equations~(\ref{eq9.1}) and (\ref{eq9.2}) result
 in the following formula for $\tau_{\rm em}$
\begin{equation}
 \tau_{\rm em} \approx ({\tau_{R}^{0}}/{\varepsilon_b}^{2})^{1/3}~,
\label{eq9.3}
\end{equation}
which is correct with the accuracy to a numerical factor,  
 which hopefully is approximately equal to unity.
 The number of oscillations within the emitted pulse is 
 \begin{equation}
 N \approx {\tau_{\rm em}}{\varepsilon_b}/{2\pi}
 \approx ({\tau_{R}^{0}} {\varepsilon_b} )^{1/3}/{2\pi}~.
\label{eq9.4}
\end{equation}
 It is remarkable that, as distinct from the radiative lifetimes of atomic
 systems, ${\tau_{\rm em}}$ is proportional to $c$ rather than to $c^3$;
 here $c$ is the speed of light. This observation implies that
 ${\tau_{\rm em}}$ is shorter than the atomic radiative lifetimes
 by the factor $(1/137)^2$, i.e., ${\tau_{\rm em}}\sim 10^{-12}$ s.

 It follows from Eq.~(\ref{eq9.3})
 that the pulse is much shorter than $\tau_{R}^0$.
 For example, if $\tau_{R}^{0}\approx 1$ ns
 and ${\varepsilon_b} \approx 10$ meV, 
 the number of oscillations equals $N \approx 4$ and the pulse duration
 is ${\tau_{\rm em}}\approx 1.5$ ps in reasonable agreement
 with the above estimate.

 Therefore, the inflating antenna shows an ultrashort radiative decay time
 $\tau_{\rm em}$ having a picosecond scale. If $\tau_{\rm em}\ll \tau$,
 all scattered photons are emitted as a single burst, 
 a short pulse having the shape of a
 train of oscillations. If $\tau_{\rm em} \gg \tau$, only a small number
 of photons are radiated in a short pulse of the duration about
 $\tau$, and the emission 
 should show a rather long tail.

 In all of the above a somewhat artificial shape of the exciting pulse,
 Eq.~(\ref{eq0}), was used. Nevertheless, the results are quite general.
 Indeed, the theory is linear in ${\bf E}_{\bf q}({\bf r}, t)$, and
 a single-side exponential pulse can be considered as a Laplace
 component of a real pulse. Therefore, the free induction signal
 generated by an arbitrary pulse can be found by integration
 of Eq.~(\ref{eq7}) over $\alpha$ with a corresponding weighting factor. 
 This integration does not change the general shape of the signal.
 The effect of smearing the sharply pointed
 pulse edge can be evaluated in a somewhat different way.
 If the termination points $t_0$ of exponential pulses
 are distributed according the Gaussian law 
 $(\gamma/\sqrt{\pi})\exp(-{\gamma}^{2}{t_0}^{2})$, the first term in
 Eq.~(\ref{eq9.0}) remains unchanged, while the second term acquires the
 factor $\exp(-{\varepsilon_b}^{2}/4{\gamma}^{2})$. If
 $\gamma \agt {\varepsilon_b}/2$, this factor results in the 
 renormalization of the coefficient $b$ by a factor of about unity. 
 It is of importance that the shape of the exciting pulse enters into the
 theory only as the initial condition, while the exotic behavior
 of the signal established above originates from the free evolution
 of the wave packet in the field of a defect.

\section{Free induction from biexcitons}
\label{sec:biexcitons}

 In this section we generalize the above results for biexcitons.
 There exist two processes that result in optical production of
 biexcitons.\cite{biex,GR73,Han,Grun} The first process is two-step absorption
 with an exciton level as a real intermediate state. In this process an
 exciton produced at the first step acts as an ``impurity.''
 All above results are applicable to this process without any serious
 changes. The second process is two-phonon absorption from the ground
 state. The theory of this process is more cumbersome than for
 impurity absorption. Nevertheless, the final results are nearly identical.

  Biexciton eigenfunctions can be written in the operator form as
\begin{equation}
 |{\bf K}j\rangle = {1\over \sqrt{2}} \int {d{\bf k}\over{(2\pi)^{3}}}
 ~\psi_{j}({\bf k})~ \psi^{\dagger}_{{\bf K}/2+{\bf k}}
 \psi^{\dagger}_{{\bf K}/2-{\bf k}}~,
\label{eq10}
\end{equation}
where $\psi^{\dagger}_{{\bf K}/2\pm {\bf k}}$ are exciton creation operators,
 and $\bf K$ is
 the center-of-mass momentum of a biexciton. Functions $\psi_{j}({\bf k})$
 are biexciton eigenfunctions in the momentum representation.
 The wave function of a biexciton wave packet 
 at positive times can be found in
 the second order of the perturbation theory in the field
 ${\bf E}_{\bf q}({\bf r}, t)$, Eq.~(\ref{eq0}).
 In the momentum representation, the wave function of the biexciton wave packet
 with a total momentum ${\bf K} = 2{\bf q}$ equals 
\begin{equation}
 \Psi_{\bf q}({\bf k}, t) =
 {2M^{2}\sqrt{V}/v\over {\omega -\varepsilon({\bf q}) + i\alpha}}
 ~A_{k0}(t)~,~~ 
 A_{k0}(t) = \sum_{j}
 { \psi_{j}({\bf k}){\bar \psi}_{j}(0) \over {2\omega -E_{j}+2i\alpha}}~
 {\rm e}^{-iE_{j}t}~.
\label{eq11}
\end{equation}
 Here $E_j$ are  energy levels of a two-exciton system, and $V$
 is the normalization volume.
 If biexcitons are excited by two light
 beams with momenta ${\bf q}_1$ and ${\bf q}_2$, function ${\bar \psi}_{j}(0)$
 in Eq.~(\ref{eq11}) should be substituted by
 $\bar{\psi}_{j}(({\bf q}_{1}-{\bf q}_{2})/2)$.

 Equation (\ref{eq11}) differs from Eq.~(\ref{eq1}) only in the coefficient
 and in the change in variables, $\omega \rightarrow \Omega = 2\omega$
 and $\alpha \rightarrow 2\alpha$. Therefore, the transformations
 that led us from  Eq.~(\ref{eq1}) to Eq.~(\ref{eq7}) can be repeated
 for biexcitons step by step. The equation for the scattering operator
 ${\cal T}_{{\bf k}'{\bf k}}(\Omega)$ depends on the interaction 
 between excitons.  The zero-radius potential provides a satisfactory
 approximation for an exciton molecule. With this potential,
 the operator ${\hat {\cal T}}(\Omega)$ for a 3D continual system has the
 form\cite{Baz}
\begin{eqnarray}
 T_{\infty} = 0, ~~~
  T(\Omega) = { {a/4\pi^{2}m}\over {1 + ia\sqrt{2m(\Omega +i0)}}  }~.
\label{eq12}
\end{eqnarray}
 Here $m$ is the exciton mass, and $\sqrt{\Omega+i0}$ has a positive
 imaginary part. Biexcitons exist when $ T(\Omega)$ has a pole, i.e.,
 for $a>0$; the scattering length $a$ equals
 $a=(2m\varepsilon_b)^{-1/2}$, where
 $\varepsilon_b$ is the biexciton binding energy. Similarly to
 ${\cal T}(\omega)$ of Eq.~(\ref{eq8}), the operator ${\cal T}(\Omega)$
 of Eq.~(\ref{eq12}) does not depend
 on momenta. Finally, $A(t)$ shows actually the same behavior as for bound
 excitons.

 One can infer from Eq.~(\ref{eq10}) that the quantum state $\Psi_{\bf q}$
 decays into two photons with momenta ${\bf K}/2 \pm {\bf k}$. 
 Intensity of the free induction signal is proportional to
 $|\Psi_{\bf q}({\bf k}, t)|^{2}\approx |\Psi_{0}(0, t)|^{2}$.
 In the asymptotic region $A(t)\propto
 \{ t + (b/2)\exp[i(\varepsilon_b t + \phi)]\}
 \exp{(-2i\varepsilon t)}$,
 and Eq.~(\ref{eq9}) describes the optically detected signal. 

 Therefore, the existence of the growing and beating modes in the 
 free induction amplitude is a common property of biexcitons and
 bound excitons.\cite{MF} Estimates for the pulse duration, $\tau_{\rm em}$, 
 derived in Sec.~\ref{sec:induction} for bound excitons, are also
 applicable to biexcitons. 

\section{Discussion}
\label{sec:Disc}

 In this paper a theory of  transients for biexcitons (bound
 excitons) is developed. It is a distinctive feature of the theory that
 the continuous spectrum and the exciton-exciton (exciton-impurity)
 interaction were consistently taken into account. Equation (\ref{eq7}) for
 the amplitude $A(t)$ and the asymptotic expansion, Eq.~(\ref{eq9.0}), 
 were found by exact solution of a well-established model. These
 equations reveal the behavior of the free induction
 signal described in Sec.~\ref{sec:induction}. The theory is reliable 
 while time $t$ is small as compared with the duration of the emitted
 pulse, $t < \tau_{\rm em}$. Under these conditions the theory
 describes the basic dynamics of exciton wave packets. The 
 larger the number of oscillations inside the emitted pulse,
 $N$, the wider the applicability region of the theory.

 It is the main restriction of the theory that the radiative decay of
 wave packets was taken into account in a
 phenomenological way. Numerical coefficients in 
 Eqs.~(\ref{eq9.3}) and (\ref{eq9.4}) for $\tau_{\rm em}$ and $N$,
 respectively, were evaluated only approximately. Actually, the values of
 these coefficients  depend
 on the number of parameters (carrier frequency of the exciting pulse,
 its spectral width, etc.). Therefore, the above restriction does not
 influence the specific predictions of the theory, but impedes
 establishing rigorous criteria of its applicability.

 The problem of the radiative decay is also
 closely related to the polariton concept. 
  It is known that in 3D polariton effects do not change the results
 critically if the longitudinal-transverse splitting, $\Delta_{lt}$,
 and the size of the coherence region are small in comparison with
 $\varepsilon_b$ and the light wavelength $\lambdabar$, respectively.
 We do not know of any experimental data on the gigantic oscillator
 strengths of bound excitons that reveal deviations from the elementary
 theory\cite{GOS} neglecting polariton effects. Apparently, the polariton
 theory\cite{HopSug} of this effect has never been applied for treating 
 experimental data. Polariton theory of the biexciton spectra was
 also devoloped,\cite{Gog,IHNH} and it was only recently that the
 detection of the polariton contribution to the giant two-photon
 biexciton absorption in CuCl has been claimed.\cite{IHNH}
 The polariton contribution to the inflating antenna theory should become
 of critical importance for large values of $t$ 
 when the size of the antenna, $d(t)$,
 approaches $\lambdabar$. Apparently, the criteria $d(t) < \lambdabar$
 and $t < \tau_{\rm em}$ impose similar restrictions on $t$.
 Therefore, a consistent theory of the radiative decay of the
 inflating antenna should be based on the polariton concept. 
 The role of polaritons depends on dimensionality,
 and in lower dimensionalities
 their effect increases.\cite{ATBC} For such systems the application
 on the polariton concept might be of special importance.

In conclusion, the coexistence of the continuous spectrum and a bound
 state results in growing and beating modes in the free
 induction signal following the exciting pulse, for both
 biexcitons and bound excitons. The duration of the free induction signal
 is controlled by the radiative decay rate and dephasing. If the first
 mechanism dominates, the signal is emitted as a short burst and the radiative
 yield is close to unity.

 \acknowledgments
I am grateful to M. D. Sturge and J. M. Worlock for suggestive discussions
 and critical
 reading the manuscript. The support of the Office of Naval Research
 under Contract No. N000149410853 is acknowledged.

\begin{figure}
 \caption{Time dependence of the amplitude $|A(t)|$
 for $\protect\alpha =
 \protect\varepsilon_{b}$, $E_{B} = 10 \protect\varepsilon_{b}$.
 Momentum $\protect{\bf k} = 0$ is at the bottom of the exciton band.
 3D:  solid lines, 2D:  dotted lines.
 (1)  $\protect\omega - \protect\varepsilon = 0 $,
 (2)  $\protect\omega - \protect\varepsilon = -~0.5  \protect\varepsilon_{b}$,
 (3)  $\protect\omega - \protect\varepsilon = -~ \protect\varepsilon_{b}$.
 For $t \protect\agt 2\protect\pi/\protect\varepsilon_{b}$ the amplitude
 shows a linear growth and undamped oscillations with a period
 $2\protect\pi/\protect\varepsilon_{b}$. }  
\label{fig1} 
\end{figure}

\begin{figure}
 \caption{ Time dependence of the amplitude $|A(t)|$ for a 3D system
 excited by a spectrally narrow pulse;
 $\protect\alpha = 0.03 \protect\varepsilon_b$,
 $E_{B}  =10\protect\varepsilon_b$. Momentum $\protect{\bf k} = 0$ 
 is at the bottom of the exciton band. Solid line:
 $\protect\omega - \protect\varepsilon = -~  \protect\varepsilon_{b}$;
 dotted line:
 $\protect\omega - \protect\varepsilon = -~0.5  \protect\varepsilon_{b}$.
}
\label{fig2}
\end{figure}


\begin{references}
\bibitem{log} There exist also dimensionality dependent logarithmic 
 corrections to the growing mode. They originate from the integration
 near ${\omega}' = \varepsilon$ in Eq.~(\ref{eq7}). We will not discuss
 these corrections here.
\bibitem{MF} There is an open question whether the phenomenon of inflating
 antenna exists in any form in dense exciton systems. If it does, it might
 be observed in Hartree-Fock data since that theory provides exciton
 amplitudes in the low-momentum region. The existence of an intimate connection
 between giant oscillator strengths and mean-field spectra has been established
 for molecular excitons (Ref. \protect\onlinecite{BRS}).
 At present one can only hypothesize
 whether the growing and oscillating four-wave-mixing signal found in
 Ref.~\protect\onlinecite{Sham} is related to inflating antenna.
\end{references}
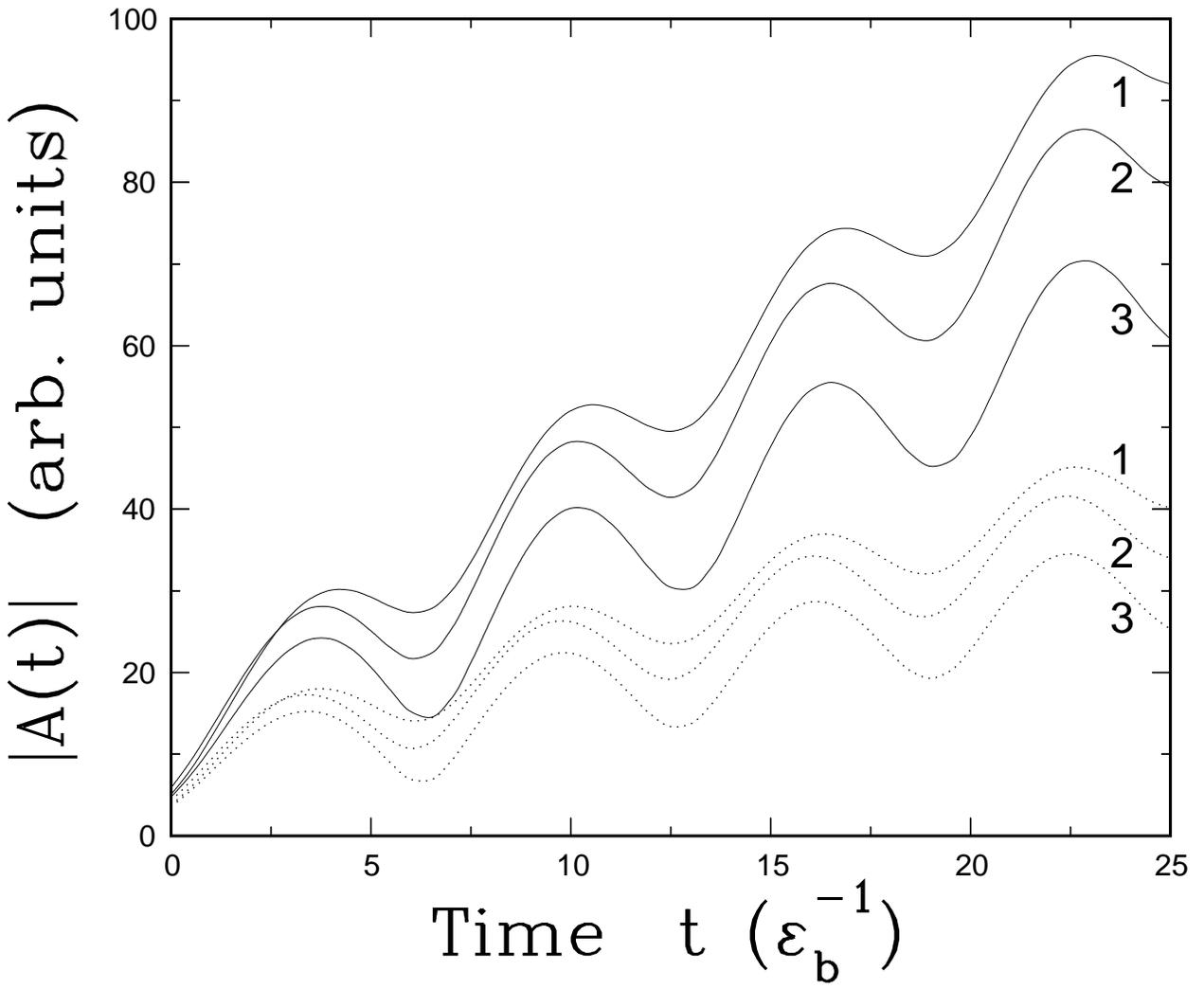
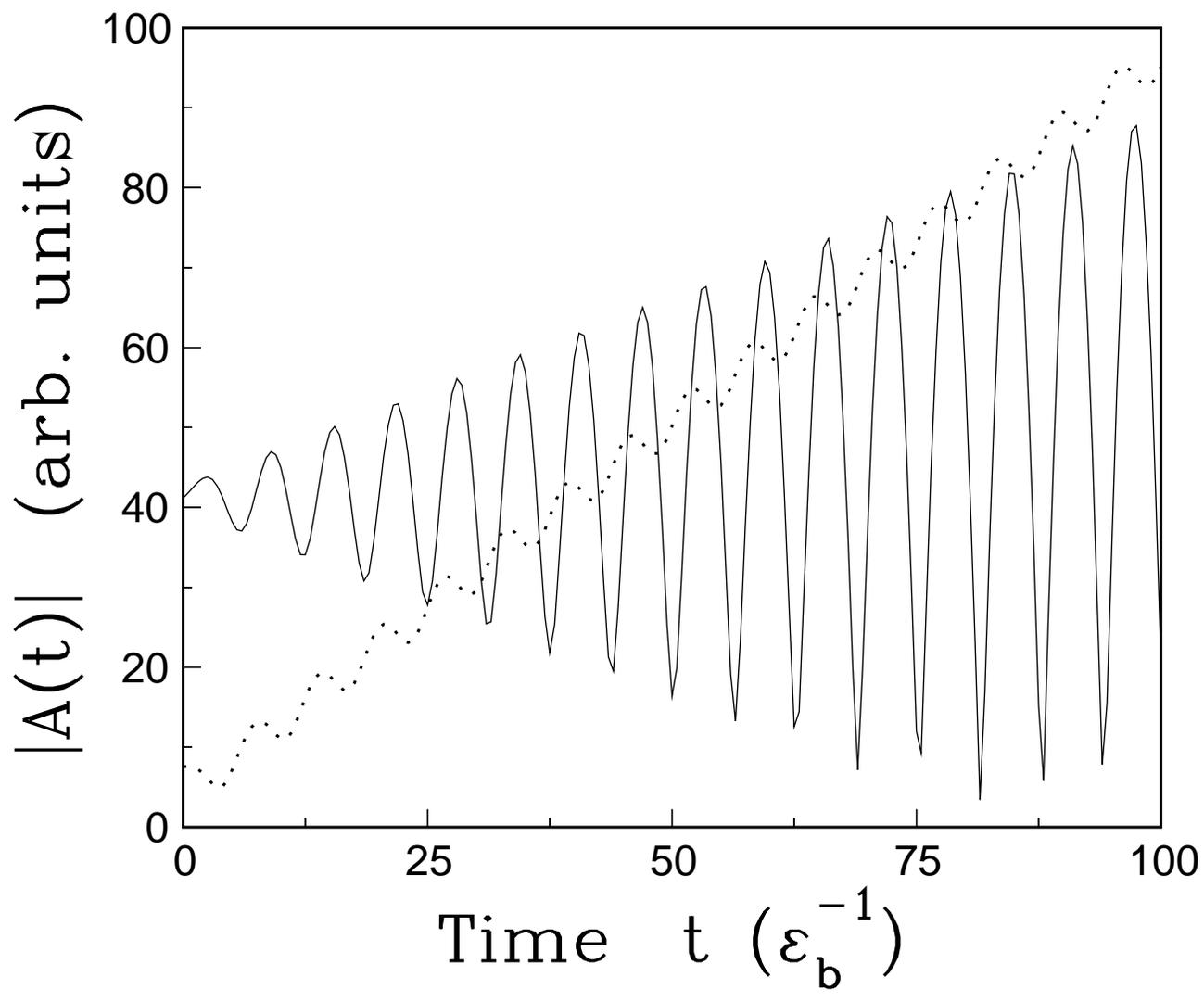
\end{document}